\documentstyle[aps,epsf]{revtex}
\begin{document}
\draft
%
%
\vspace*{-20mm}
\leftline{\epsfbox{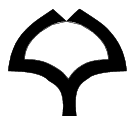}}
\vspace*{-10mm}
{\baselineskip-4pt
\font\yitp=cmmib10 scaled\magstep2
\font\elevenmib=cmmib10 scaled\magstep1  \skewchar\elevenmib='177
\leftline{\baselineskip20pt
\hspace{12mm} 
\vbox to0pt
   { {\yitp\hbox{Osaka \hspace{1.5mm} University} }
     {\large\sl\hbox{{Theoretical Astrophysics}} }\vss}}

%
%
{\baselineskip0pt
\rightline{\large\baselineskip14pt\rm\vbox
        to20pt{\hbox{November 2001}
               \hbox{OU-TAP-168}
\vss}}
}
\vskip20mm

\begin{center}
{\Large\bf   Cosmic Inversion\\
~~\\
---Reconstructing primordial spectrum from CMB anisotropy---}
\end{center}
\vspace{3mm}

\begin{center}
{\large Makoto Matsumiya,\footnote{E-mail:matumiya@vega.ess.sci.osaka-u.ac.jp}
  Misao Sasaki\footnote{E-mail:misao@vega.ess.sci.osaka-u.ac.jp}
  and \,Jun'ichi Yokoyama\footnote{E-mail:yokoyama@vega.ess.sci.osaka-u.ac.jp}}
\vskip 1mm
\sl{Department of Earth and Space Science,
Graduate School of Science,\\
Osaka University, Toyonaka 560-0043, Japan}
\end{center}

\begin{abstract}
We investigate the possibility of reconstructing the initial
spectrum of density fluctuations from the cosmic microwave
background (CMB) anisotropy. As a first step toward this program,
we consider a spatially flat, CDM dominated universe. In this case,
it is shown that, with a good accuracy, the initial spectrum satisfies
a first order differential equation with the source determined by
the CMB angular correlation function. The equation is found to contain
singularities arising from zeros of the acoustic oscillations in the
transfer functions. Nevertheless, we find these singularities are not
fatal, and the equation can be solved nicely.
We test our method by considering simple analytic forms for the
transfer functions. We find the initial spectrum is reproduced
within 5\% accuracy even for a spectrum that has a sharp spike.
\end{abstract}


\section{Introduction}

Determination of the primordial spectrum of the density fluctuations
is one of the most important issues in modern cosmology.
The cosmic microwave background (CMB) anisotropy
provides us with a great deal of information of the primordial fluctuations,
and it is considered to be a powerful tool for studying
the early universe \cite{WSS1}.
This is because the physical processes that determine the CMB anisotropy
are described by linear perturbation theory, and they have been well
understood \cite{PY,HS}.

Although the recent anisotropy observations are consistent with
a flat universe with a scale-invariant initial power spectrum,
this does not exclude the possibilities of other models.
In particular, we do not know how much the shape of
the initial power spectrum is constrained.
In most of previous investigations, when cosmological model parameters
are estimated from the observational data by likelihood analysis,
the initial spectrum is assumed to have a power-law shape \cite{power}.
It is true that a conventional slow-roll inflation model \cite{REV}, that
has now become a `standard model', gives a power-law spectrum
which is almost scale-invariant\cite{PER}.
However, when analyzing the observed CMB anisotropy,
it is much more desirable, and probably much healthier, to constrain
the initial spectrum solely by observed data without
any theoretical prejudices.
For example, even within the context of inflationary cosmology,
a variety of generation mechanisms for
non-scale-invariant perturbations have been proposed \cite{BSI}.
In this connection, recently several authors have discussed extraction
of non-power law features from the CMB observations \cite{SOU},
where the initial spectrum is allowed to have a piece-wise power-law
shape.

In this paper, we approach this issue in an entirely different way,
namely, by formulating an inverse problem as faithful as possible.
Such an approach will be eventually needed if we seriously want
to constrain the initial power spectrum solely from observations of
the CMB anisotropy.
An approach to this inversion problem has been discussed
recently \cite{BER}.

As a first step, we consider a simple situation in which the
transfer functions that relate the input power spectrum $P(k)$
of the gravitational potential to the output CMB angular
correlation function $C(\theta)$ are given analytically.
This is certainly a toy model. However, it has almost all the essential
features a realistic model would have.
In particular, unlike \cite{BER}, our model takes account
of not only the Sachs-Wolfe (SW) effect\cite{SW} but also
the Doppler effect . The latter, which gives rise to
zero points in the transfer functions, is the main cause
of the difficulty in this inversion problem.

The advantage of adopting this simple situation is that
our method of inversion which we shall develop below may be
easily tested at various stages of calculations.
Since our primary concern here is to formulate the inversion problem,
we fix the cosmological parameters and do not study the dependence of
$P(k)$ upon them.

The paper is organized as follows.
In Sec.~II, we give the basic equations
that relate the primordial spectrum $P(k)$ with
the angular correlation function $C(\theta)$.
In Sec.~III, under some reasonable assumptions,
we derive a differential equation for $P(k)$
and develop a method to solve it.
Then we test our method by applying it to
several spectral shapes. We find our method is
applicable even in the case of a spectrum with
a sharp spike.

\section{Basic Equations}
The angular correlation function of the CMB, $C(\theta)$ is defined as
\begin{equation}
C(\theta)=\langle \Theta(\bbox{\gamma}_1)
\Theta(\bbox{\gamma}_2) \rangle,
\hspace{1.0cm}\cos \theta = \bbox{\gamma}_1 \cdot
\bbox{\gamma}_2
\end{equation}
where $\Theta(\bbox{\gamma})=
(\Delta T/\bar{T})(\bbox{\gamma})$ is
the temperature fluctuation in the direction $\bbox{\gamma}$
and the average is taken over all angles and all spatial
positions. $C(\theta)$ is expanded in the Legendre polynomials
and related to the angular power spectrum, $C_l$, as
\begin{equation}
C(\theta) = \sum_l \frac{2l+1}{4\pi}C_l P_l(\cos \theta).
\label{Cell}
\end{equation}
$C(\theta)$ or $C_l$ are the fundamental observables of the CMB
anisotropy. Our purpose is to reconstruct the initial power spectrum
from the above observables.
We calculate in the Newtonian gauge and follow the
notation of \cite{HS}.

Each Fourier mode of temperature perturbations,
$\Theta(\eta,\bbox{k})$, obeys
the Boltzmann equation \cite{KS},
\begin{equation}
\dot{\Theta}+ik\mu (\Theta + \Psi)=-\dot{\Phi}
+\dot{\tau}[\Theta_0-\Theta-\frac{1}{10}\Theta_2 P_2(\mu)-i\mu V_b],
\label{Boltzmann}
\end{equation}
where the dot denotes a derivative with respect to the conformal time
$\eta$, $k$ is the comoving wave number,
$\mu= k^{-1} \bbox{k}\cdot \bbox{\gamma}$,
$\dot{\tau}$ is the differential
Thomson optical depth and $V_b$ is the bulk velocity of baryons.
$\Psi$ and $\Phi$ are the gauge-invariant Newtonian potential
and spatial curvature perturbation, respectively\cite{KS}.
Here, we decompose $\Theta(\eta,k,\mu)$ into the multipole moments,
\begin{equation}
\Theta(\eta,k,\mu)=\sum_l (-i)^l \Theta_l(\eta,k) P_l(\mu),
\end{equation}
where
$\Theta_l(\eta,k)$
is the $l$-th multipole moment of $\Theta(\eta,k,\mu)$.
Integrating Eq.~(\ref{Boltzmann}), we obtain
\begin{equation}
(\Theta+\Psi)(\eta_0,k,\mu)=\int^{\eta_0}_0 \,
\left\{[\Theta_0+\Psi-i\mu V_b]
  {\cal V}(\eta)+(\dot{\Psi}-\dot{\Phi})e^{-\tau(\eta)}\right\}
                        e^{ik\mu(\eta-\eta_0)}d \eta,
\end{equation}
where $\eta_0$ is the conformal time today and
${\cal V}(\eta)$ is the visibility function given by
\begin{eqnarray}
{\cal V}(\eta) =  \dot{\tau}(\eta)e^{-\tau(\eta)}\,;
\quad
\tau(\eta) =  \int^{\eta_0}_{\eta} \dot{\tau}(\eta^{\prime})
                 d \eta^{\prime}.
\end{eqnarray}
We have neglected the quadrupole term on the right hand side
of Eq.~(\ref{Boltzmann}) since its contribution is
negligible in the tight coupling approximation.
The visibility function has a sharp peak around the
last scattering time $\eta_*$ so that we assume that recombination occurs
instantaneously at $\eta=\eta_*$.
Then, the multipole moments of each $k$ mode is approximately given by
\begin{eqnarray}
 \Theta_l(\eta_0,k) & = & (\Theta_0+\Psi)(\eta_{*},k)(2l+1)j_l(kd)
+\Theta_{1}(\eta_*,k)(2l+1)j^{\prime}_l(kd) \nonumber\\
& &+(2l+1)\int^{\eta_0}_{\eta_*}
d\eta^\prime
\frac{\partial}{\partial \eta^{\prime}}
(\Psi(\eta^\prime,k)
-\Phi(\eta^\prime,k))j_l(k\eta_0-k\eta^\prime),
\label{thetal}
\end{eqnarray}
where $d=\eta_0-\eta_*$ is a conformal distance from the present epoch
to the last scattering surface (LSS).
Equation~(\ref{thetal})
shows that there are four sources for the temperature anisotropy,
namely, the intrinsic temperature variation,
the Sachs-Wolfe (SW) effect which is caused by the static gravitational
potential at the last scattering surface and is dominant on large scales,
the Doppler effect due to the fluid bulk velocity, and
the integrated Sachs-Wolfe (ISW) effect
due to the time variation of the metric perturbation after decoupling.
If the recombination occurs when the universe is matter-dominated,
$\Psi$ and $\Phi$ remain constant in time
after decoupling until a stage at which either the curvature term
or the cosmological constant term becomes significant.
In particular, in the Einstein-de Sitter universe,
they remain constant until today, hence the ISW effect is absent.

Conventionally, $C_l$ is expressed as
\begin{equation}
\frac{2l+1}{4\pi}C_l=\frac{1}{2\pi^2} \int^{\infty}_{0} dk \, k^2
\frac{|\Theta_l(\eta_0,k)|^2}{2l+1}\,.
\label{cl}
\end{equation}
From Eqs.~(\ref{thetal}) and (\ref{cl}), we find
\begin{equation}
C_l = \frac{2}{\pi} \int^\infty_0 dk\,k^2 \Bigl[
(\Theta_0+\Psi)(\eta_*)j_l(kd)+\Theta_1(\eta_*)
j^{\prime}_l(kd) \Bigr]^2.
\label{clapp}
\end{equation}
The angular correlation function is calculated from Eq.~(\ref{clapp}).
Here, we define $r$ by
\begin{equation}
r=2d\sin\frac{\theta}{2}\,.
\label{rdef}
\end{equation}
This is the spatial distance between two points
on the last scattering surface which are observed
with the angular separation $\theta$.
Since the thickness of the LSS is neglected in
Eq.~(\ref{thetal}),  there is a
one-to-one correspondence between the observed
temperature anisotropy and the perturbation variables on the LSS.
Using the relation,
\begin{equation}
\sum^{\infty}_{l=0}(2l+1)P_l(\cos \theta)j^{2}_l(kd)
= \frac{\sin kr}{kr}\,,
\end{equation}
the angular correlation function $C(r)$ is given by
\begin{eqnarray}
C(r) & = & \frac{1}{2\pi^2} \int^\infty_0 dk\,k^2
\Biggl[|\Theta_0+\Psi|^2 +(\Theta_0+\Psi)
\Theta_1 \frac{r}{kd}
\frac{\partial}{\partial r} \nonumber\\
& & \hspace{1.0cm} +\frac{|\Theta_1|^2}{k^2}\left\{\left(-\frac{1}{r}
+\frac{r}{4d^2} \right)\frac{\partial}{\partial r} +
\left(\frac{r}{2d} \right)^2 \frac{\partial^2}{\partial r^2}
\right\}
\Biggr]\frac{\sin kr}{kr} \nonumber \\
& = &\frac{1}{2\pi^2} \int^\infty_0 dk\,k^2
\Biggl[|\Theta_0+\Psi|^2 \frac{\sin kr}{kr}
+\frac{|\Theta_1|^2}{k^2 r^2}\left(\frac{\sin kr}{kr}
-\cos kr\right) \nonumber \\
& &\hspace*{2.0cm}+\frac{(\Theta_0+\Psi)\Theta_1}{kd}
 \left(-\frac{\sin kr}{kr}
+\cos kr \right) \nonumber \\
& &\hspace*{2.0cm}+\frac{|\Theta_1|^2}{4 k^2 d^2}
\left(\frac{\sin kr}{kr}-\cos kr -kr \sin kr \right)
 \Biggr].
\label{cr}
\end{eqnarray}

In the linear perturbation theory, different $k$ modes do not couple
with each other but evolve independently. The initial
condition for each $k$ mode is directly reflected in the
perturbation variables through transfer functions.
The variables, $\Theta_0$, $\Theta_1$ and $\Psi$,
in the integrand of Eq.~(\ref{cr}) are therefore linearly related
to the initial condition and
we can generally express the relation between $C(r)$
and the initial spectrum as
\begin{equation}
C(r)=\int^\infty_0 K(k,r) P(k) dk,
\label{integraleq}
\end{equation}
where we normalize the initial condition in terms of the curvature
perturbation $\Phi(\eta=0,k)$
and define $P(k) \equiv \langle |\Phi(0,k)|^2 \rangle$.
In this case, introducing the transfer functions $f(k)$ and $g(k)$
defined by
\begin{eqnarray}
(\Theta_0+\Psi)(\eta_*,k) & = & f(\eta_*,k)\Phi(0,k),
\nonumber\\
\Theta_1(\eta_*,k) & = & g(\eta_*,k)\Phi(0,k),
\end{eqnarray}
$K(k,r)$ can be written as
\begin{equation}
K(k,r)=\frac{k^2}{2 \pi^2 r^3}\left(\sum^2_{i=0} \alpha_{2i}(k) r^{2i}
\sin kr
       +\sum^1_{i=0}\alpha_{2i+1}(k)r^{2i+1}\cos kr\right),
\label{kernel}
\end{equation}
where $\alpha_i(k)$ are given by
\begin{eqnarray}
\alpha_0(k) & = & \frac{g^2(k)}{k^3},\nonumber \\
\alpha_1(k) & = & -\frac{g^2(k)}{k^2}, \nonumber \\
\alpha_2(k) & = & \frac{f^2(k)}{k}-\frac{f(k)g(k)}{k^2 d}
               +\frac{g^2(k)}{4k^3 d^2}, \nonumber \\
\alpha_3(k) & = & \frac{f(k)g(k)}{kd}-\frac{g^2(k)}{4k^2d^2},\nonumber \\
\alpha_4(k) & = & -\frac{g^2(k)}{4kd^2}.
\end{eqnarray}

Our purpose is to solve the integral equation (\ref{integraleq})
for $P(k)$ with the kernel given by Eq.~(\ref{kernel}).
When there is only the monopole term,
{\it i.e.}, $\Theta_1=0$, Eq.~(\ref{integraleq}) takes the familiar form,
\begin{equation}
C(r)=\frac{1}{2 \pi^2} \int^{\infty}_0 dk\, k^2 \, f^2(k) P(k)
\frac{\sin kr}{kr}.
\end{equation}
In this case, using the Fourier sine formula, we obtain
\begin{equation}
P(k)= \frac{4 \pi}{f^2(k) k} \int^{\infty}_0 dr\,r\,C(r) \sin kr.
\end{equation}
It must be noted that $P(k)$ has singularities on small scales since
$f(k)$ and $g(k)$ are oscillatory functions reflecting the
acoustic oscillations of the density perturbations.
These singularities are inevitable as long as we take this approach
and the one-to-one correspondence
between $C(r)$ and $P(k)$ breaks down at the singularities.
However, we shall find in the next
section that there is a way to resolve this difficulty.

\section{Inversion Method}

In the above discussion, we have tacitly assumed that  $r$ runs from
zero to infinity.  In reality, however, $r$ is bounded
in the finite range $0 \leq r \leq 2d$. Furthermore,
it is observationally impossible to determine $C(r)$ on large scales
due to the statistical ambiguity, {\it i.e.}, the cosmic variance.
However, the scales which we are
interested in are $r \ll d$ and it is expected
that modes with $k \geq 1/d$
have little effect on these scales.
We therefore neglect the terms proportional to $1/d$ and $1/d^2$
in Eq.~(\ref{cr}). In this limit, we have
\begin{equation}
C(r)=\frac{1}{2\pi^2} \int^{\infty}_{0} dk \frac{P(k)}{k r^3} \left\{
      F(k)k^2 r^2 \sin(kr)-G(k)kr \cos kr+G(k) \sin kr \right\},
\label{crapp}
\end{equation}
where we have replaced $f^2(k)$ and $g^2(k)$ with $F(k)$ and $G(k)$,
respectively, for notational simplicity.
Multiplying by $r^3$ and integrating by parts
twice, we may re-express Eq.~(\ref{crapp}) as
\begin{equation}
r^3 C(r) = \frac{1}{2 \pi^2} \int^{\infty}_{0} dk \, \left\{
           -\frac{\partial^2}{\partial k^2}(kF(k)P(k))
           +\frac{\partial}{\partial k}(G(k)P(k))+\frac{1}{k}G(k)P(k)
           \right\}\sin kr\,,
\end{equation}
where we have assumed that the boundary terms
vanish and the integral converges.
Employing the Fourier sine formula, we obtain
\begin{eqnarray}
-kF(k)P^{\prime \prime}(k)&& +
 (-2F-2kF^{\prime}+G)P^{\prime}(k)
\nonumber\\
+&&
(G^{\prime}-2F^{\prime}-kF^{\prime \prime}+G/k)P(k)
=4 \pi
\int^{\infty}_{0} r^3 C(r) \sin kr dr.
\label{eq2diff}
\end{eqnarray}
This is a second-order ordinary differential equation
in which the source term carries the information of $C(r)$.
It is of course necessary to
fix the boundary conditions for $P(k)$ in order to
solve Eq.~(\ref{eq2diff}). However,
we have no means to determine these conditions.
The only knowledge we have is that
the boundary conditions at $k=0$ and $k=\infty$ are restricted
from the convergence of the integral in Eq.~(\ref{crapp}).

In order to avoid this difficulty, we consider the
reduction of the order of the differential equation.
If we take some appropriate combination of $C(r)$ and
its derivative, we can derive a differential equation of lower order.
The simplest combination is given by
\begin{equation}
\tilde C(r)\equiv
3rC(r)+r^2C^{\prime}(r)=\frac{1}{2\pi^2}\int^{\infty}_{0} dk \,
P(k)\left\{F(k)k^2 r \cos kr +(2F(k)+G(k))k \sin kr \right\}.
\label{cr+crd}
\end{equation}
Note that the power of $r$ in the integrand is reduced. No linear
combination can reduce the order of the differential equation lower than
$\tilde C(r)$.
Integrating by parts, we obtain
\begin{equation}
-F k^2 P^{\prime}+(-F^{\prime} k+G)kP=4 \pi \int^{\infty}_{0}
\tilde{C}(r) \sin kr dr\,.
\label{eqdif}
\end{equation}
This is a first order differential equation. We discuss a method
for solving this equation in the rest of this section.

We now give the expressions for $f(k)$ and $g(k)$ explicitly
in order to examine the properties of Eq. (\ref{eqdif}), and to
verify if thus obtained solution correctly reproduces the original
spectrum. For a given model,
$f(k)$ and $g(k)$ are determined by the coupled Einstein-fluid equations
which are to be solved numerically.
However, in order to understand the property
of Eq.~(\ref{eqdif})
and to establish a method which we can apply to general cases,
we start with a toy model in which $f(k)$ and $g(k)$
are given analytically.

Before recombination, photons are tightly coupled with baryons
through the Thomson scattering. With the tight coupling approximation,
we can expand the Boltzmann equation for photons together with the
continuity and Euler equations for baryons in the
Thomson scattering time \cite{PY}.
To the leading order, the dynamics of the photon-baryon fluid is
described by the following simple
equation for each Fourier mode \cite{HS},
\begin{equation}
\ddot{\Theta}_{0}+\frac{\dot{a}}{a} \frac{R}{1+R} \dot{\Theta}_{0}+k^2
 c^{2}_{s}\Theta_{0} = {\cal F}(\eta),
\label{monopole}
\end{equation}
where $c_{s}$ is the sound speed of the photon-baryon fluid,
$R \equiv 3\rho_{b}/4\rho_{\gamma}$, and
${\cal F}(\eta)$ is the driving term
due to the gravitational potential,
\begin{equation}
{\cal F}(\eta)=-\ddot{\Phi}-\frac{\dot{a}}{a} \frac{R}{1+R} \dot{\Psi}
-\frac{k^2}{3}\Psi\,.
\end{equation}
The dipole moment is obtained from the continuity equation,
\begin{equation}
k\Theta_{1} = -3(\dot{\Theta}_{0}+\dot{\Psi})\,.
\label{dipole}
\end{equation}
Note that Eqs.~(\ref{monopole}) and (\ref{dipole}) are applicable not
only to the Einstein-de Sitter universe but also to other cosmological
models.

Here we solve Eqs.~(\ref{monopole}) and (\ref{dipole})
under the following assumptions:
\begin{list}{}{}
\item[(i)] The gravitational potential fluctuation $\Psi(\eta)$
and the spatial curvature fluctuation $\Phi(\eta)$ are
always independent of time, {\it i.e.}, $\dot{\Psi}=\dot{\Phi}=0$, and
$\Psi=-\Phi$.
\item[(ii)] The baryon denisty is negligible, {\it i.e.}, $R=0$.
\item[(iii)] The perturbation is adiabatic.
\end{list}
 The assumption (i) is valid only on large and
intermediate scales, but fails on small scales ({\it i.e.}, on scales
much smaller than the Hubble horizon scale at decoupling)
since the acoustic oscillations tend to destroy the gravitational potential
on such scales.  The assumption (ii) is, of course, quite unrealistic,
while (iii) is a natural consequence of inflationary cosmology.
However, these assumptions do not alter the essential features
of the CMB anisotropy spectrum.
The difficulty of solving Eq.~(\ref{eqdif})
is caused by the acoustic oscillations of temperature fluctuations
which give rise to zero points in the transfer functions $f$ and $g$,
and this is a common feature of the CMB anisotropy
regardless of cosmological models.
Furthermore, we do not take account of
the diffusion damping \cite{SI} which becomes significant
on small scales, since this effect do not significantly
change the structure of Eq.~(\ref{eqdif}) either.

We solve Eqs.~(\ref{monopole}) and (\ref{dipole}) in this toy model.
The solutions are given by
\begin{eqnarray}
[\Theta_{0}+\Psi](\eta) &=& \frac{1}{3}\Phi(0) \cos(kc_{s}\eta),
\nonumber \\
\Theta_{1}(\eta) &=& c_{s} \Phi(0) \sin(kc_{s}\eta),
\label{sol}
\end{eqnarray}
where $c_{s}=1/\sqrt{3}$ and we take the adiabatic
initial condition as $\Theta_0(0)=\Psi(0)/3=-\Phi(0)/3$ and
$\Theta_1(0)=0$.
{}From Eq.~(\ref{eqdif}), we obtain
\begin{equation}
-\frac{1}{9}\cos^2 kr_* k^2 P^{\prime}+\left(\frac{2}{9}kr_* \cos kr_*
\sin kr_* +\frac{1}{3}\sin^2 kr_* \right) k P=S(k)\,,
\label{diff}
\end{equation}
where $r_*=c_s\eta_*$ and
\begin{eqnarray}
S(k)=4 \pi\int^{\infty}_{0} \tilde{C}(r) \sin kr\, dr \,.
\label{Sdef}
\end{eqnarray}

Now let us describe our method.
When the data of $\tilde{C}(r)$ are given, $S(k)$ is computed from the
Fourier sine transform of it. In actual situations,
$\tilde{C}(r)$ is to be obtained by observation. However, here
we use $\tilde{C}(r)$ obtained from Eq.~(\ref{cr}) by giving an original
spectrum by hand.
Then we solve Eq.~(\ref{diff})
and compare the solution with the original spectrum.
Note that we do not use Eq.~(\ref{crapp}) for the evaluation of
$\tilde C(r)$ since
it is an approximate formula obtained by taking the limit
$d \to \infty$. This is because our purpose is to examine
the accuracy of our method which uses several approximations.
For the distance to the LSS, we choose $d=100\,r_*$.

The theoretical expression for $\tilde{C}(r)$
includes contributions from $C_l$ of all $\ell$;
$0 \leq l \leq \infty$. In reality, the monopole and dipole
contributions of the anisotropy cannot be observed and the higher
multipole moments are limited by the angular resolution of observation.
Therefore, we subtract these contributions from  $\tilde{C}(r)$ and
define the following function instead of Eq.~(\ref{Cell}),
\begin{equation}
{C}_{obs}(r)=\sum^{l_0}_{l=2}\frac{2l+1}{4\pi}
C_{l}P_{l}\left(1-\frac{r^2}{2 d^2}\right),
\end{equation}
where $l_0$ is an upper limit of $l$ and
we take $l_0=2000$ in the following calculation.
We denote the corresponding $\tilde C(r)$ by $\tilde C_{obs}(r)$.
Since the $l=0$, $1$ terms contribute to $\tilde C(r)$ mainly
at large $r$,
the source term is not expected to be modified significantly by neglecting
these terms. As for the contributions of $l\geq l_0$,
they are negligible since $C_l$ decreases exponentially for large $l$.
In Fig.~\ref{corr}, we show $\tilde{C}(r)$ and $\tilde{C}_{obs}(r)$
for a power law spectrum with a damping factor,
$P(k)= (kd)^{-3}\exp(-k/k_0)$.
In this case, $\tilde{C}(r)$ can be calculated analytically and
we can estimate the effect of using the finite data set of $C_l$.
We also plot $\tilde{C}_{app}(r)$ which is given
by cutting off the high multipole contributions of $l\geq l_0$
from $\tilde{C}(r)$.
The relative error between $\tilde{C}_{app}(r)$
and the analytic $\tilde{C}(r)$ is below 2\,\%,
which justifies the neglect of the high multipole contributions.
On the other hand, $\tilde{C}_{obs}(r)$ differs from $\tilde{C}(r)$
substantially. However, since we are
interested in scales much smaller than $d$,
this subtraction of $l=0,1$ modes is expected to be (and in fact found
to be) harmless.

\begin{figure}[h]
\begin{center}
\leavevmode
\epsfxsize=7.0cm
\epsfbox{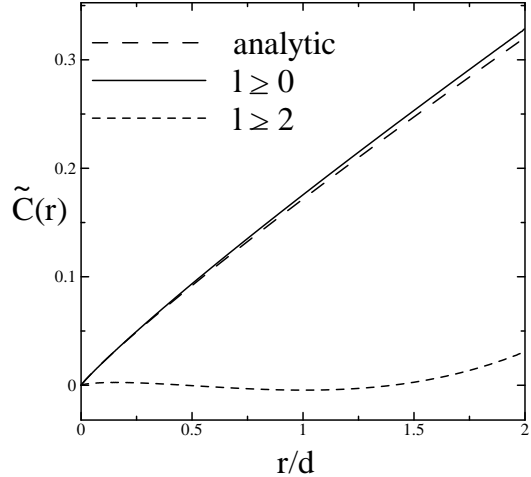}
\caption{$\tilde{C}(r)$ $(0\leq l <\infty$),
$\tilde{C}_{app}(r)$ $(0\leq l\leq2000)$
and $\tilde{C}_{obs}(r)$ $(2 \leq l \leq 2000)$
for the spectrum $P(k)= (kd)^{-3}\exp(-k/k_0)$.
We set $k_0 d=500$. The relative
error between $\tilde{C}_{app}(r)$ and $\tilde{C}(r)$
is less than 2 \%. }\label{corr}
\end{center}
\end{figure}

\begin{figure}[h]
\begin{center}
\leavevmode
\epsfxsize=7.0cm
\epsfbox{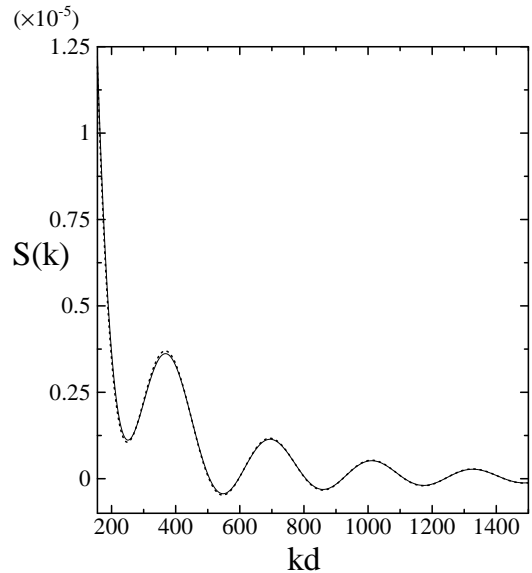}
\caption{The exact source term (solid line) and the source term
calculated from the
Fourier transform of $\tilde{C}_{obs}(r) $ (dotted line)
for the spectrum same as Fig.~\protect\ref{corr}.
The relative error, which results from the small scale
 approximation, $r/d \ll 1$ and the finiteness of the $C_l$ data,
is below 5\,\%.}
\label{sk}
\end{center}
\end{figure}

For a given spectrum, the exact $S(k)$
can be evaluated by simply substituting the original $P(k)$
into the left-hand side of Eq.~(\ref{diff}).
Since we are interested in the inversion problem,
we need to know how accurately the source term $S(k)$
can be calculated from $\tilde{C}_{obs}(r)$. In particular,
we must reproduce the exact one with a good accuracy
on scales $l \gtrsim 100$.
We therefore examine the accuracy of the calculated $S(k)$
by comparing it with the exact one.

Before we make this comparison, however, there is
one more problem to be resolved. It is the problem of
the range of definition of $r$.
The Fourier transform of $\tilde{C}(r)$ cannot be performed in the exact sense
since $r$ given by Eq.~(\ref{rdef}) is not defined for $r>2d$.
To avoid this difficulty, we cut off
$\tilde{C}_{obs}(r)$ by a smooth function which damps out
steeply at some scale $r \gtrsim 0.01d$ which is much smaller than
$2d$ but large enough to reproduce the spectrum over a sufficiently wide
range. Although this lowers the amplitude of large $k$-modes,
the function obtained by the Fourier transform has little deviation
from the exact $S(k)$ on scales of our interest.
Both the exact and calculated $S(k)$
are shown in Fig.~\ref{sk}.
Since we take $l_0=2000$, the region where the calculated source term
matches the exact one is below $kd=l \leq 2000$.

As mentioned before, Eq.~(\ref{diff}) has
singularities at $kr_*=(n+1/2)\pi$ which cause
difficulties when we solve it numerically.
We can, however, determine the values of $P(k)$ at these
singularities if we assume that the derivative of
$P(k)$ is finite. Then the first term on the left-hand side
of Eq.~(\ref{diff}) vanishes at $kr_*=(n+\frac{1}{2})\pi$, and
the values of $P(k)$ at these singularities are given by
\begin{equation}
P \left[k= \left(n+\frac{1}{2}\right) \frac{\pi}{r_*} \right] =
\frac{3r_*}
{\left(n+\frac{1}{2}\right)\pi}
S\left[k= \left(n+\frac{1}{2}\right) \frac{\pi}{r_*} \right].
\end{equation}
Once these values are given, we can solve Eq.~(\ref{diff}) by expanding
it around the singularities.
We search the true solution which connect the adjacent
singularities using the shooting method.
We solve the equation until the
5th singularity, $kd=450 \pi$, for the following two cases
of the original spectra:
\begin{eqnarray}
&&P(k)=\frac{(kd)^{-3}}{1+(k/k_s)^p}\exp(-k/k_0),
\label{double}
\\
&&P(k)=(kd)^{-3}
\left[1+A \exp \left\{-\frac{(kd-k_p d)^2}{\sigma} \right \}\right]^{s}
\exp(-k/k_0)\,;\quad s=\pm1\,.
\label{peaked}
\end{eqnarray}
The first one is a double power-law spectrum and the
second is a single power-law spectrum with a spiky structure,
either with a peak ($s=+1$) or a dip ($s=-1$).
The results for the choice of the parameters
as $p=2$, $A = 10$, $k_p d=600$, $\sigma =10$ and $k_0 d=1000$
are shown in Fig.~\ref{pk}.
We find our
method reproduces the original spectra with a good accuracy.
In particular, even if the spectrum has a sharp peak or a dip,
we can resolve such a local structure using this method.
The numerical solution diverges as it approaches the singularities
(indicated by the triangles),
but the relative error except for the regions close to the
singularities is below 5\,\%.

For comparison, in Fig.~\ref{clspectrum}, we also show the corresponding
CMB angular power spectra calculated by
using the transfer functions given by Eq.~(\ref{sol}).
In the case of the spectrum with a peak, one can clearly see an
enhancement of $C_l$ at $l\sim600$, though the peak height is
suppressed by a factor of $\sim 4$ as compared to the original peak
in the primordial spectrum. As for the spectrum with a dip,
the structure is even more suppressed and leaves only a small imprint
in the angular spectrum.
In a realistic case, a small structure may be present
in the CMB spectrum as a consequence of an irregular feature
in the primordial spectrum. The above result indicates that
our method is capable of reproducing such a feature from
the CMB spectrum.

\begin{figure}
\begin{center}
\leavevmode
\hspace*{-1cm}
\epsfxsize=5.5cm
\epsfbox{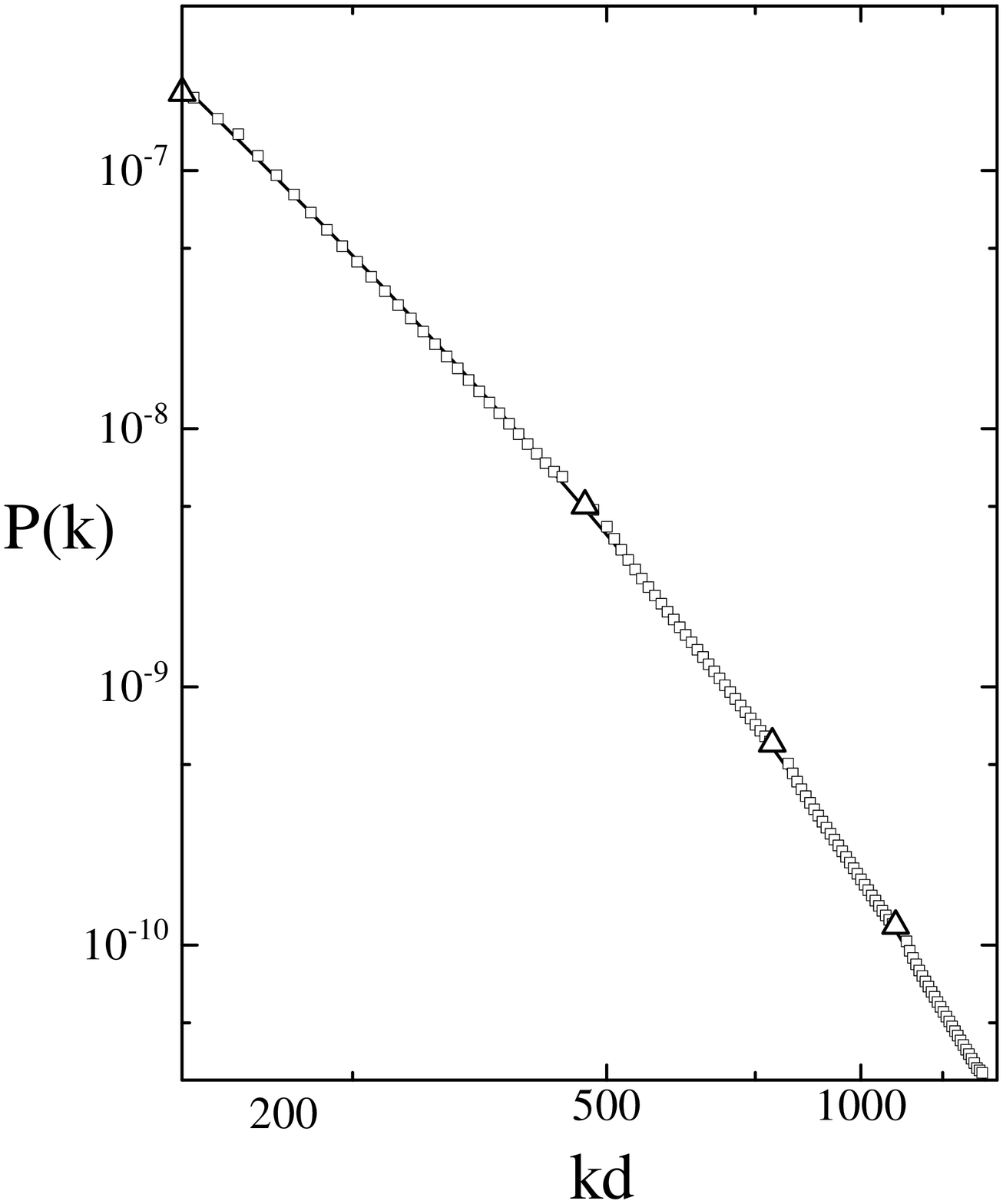}
\hspace*{0.4cm}
\epsfxsize=5.5cm
\epsfbox{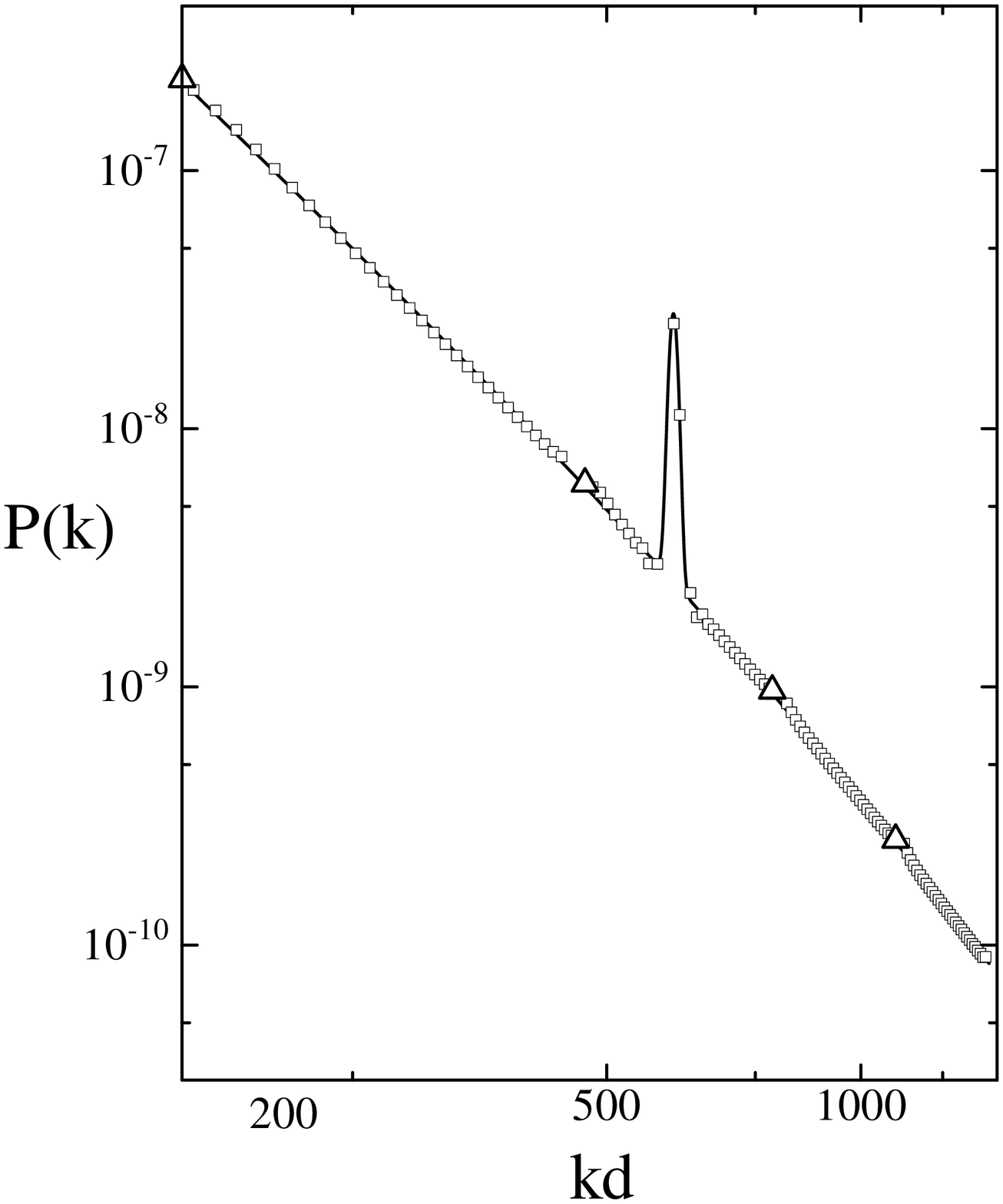}
\hspace*{0.4cm}
\epsfxsize=5.5cm
\epsfbox{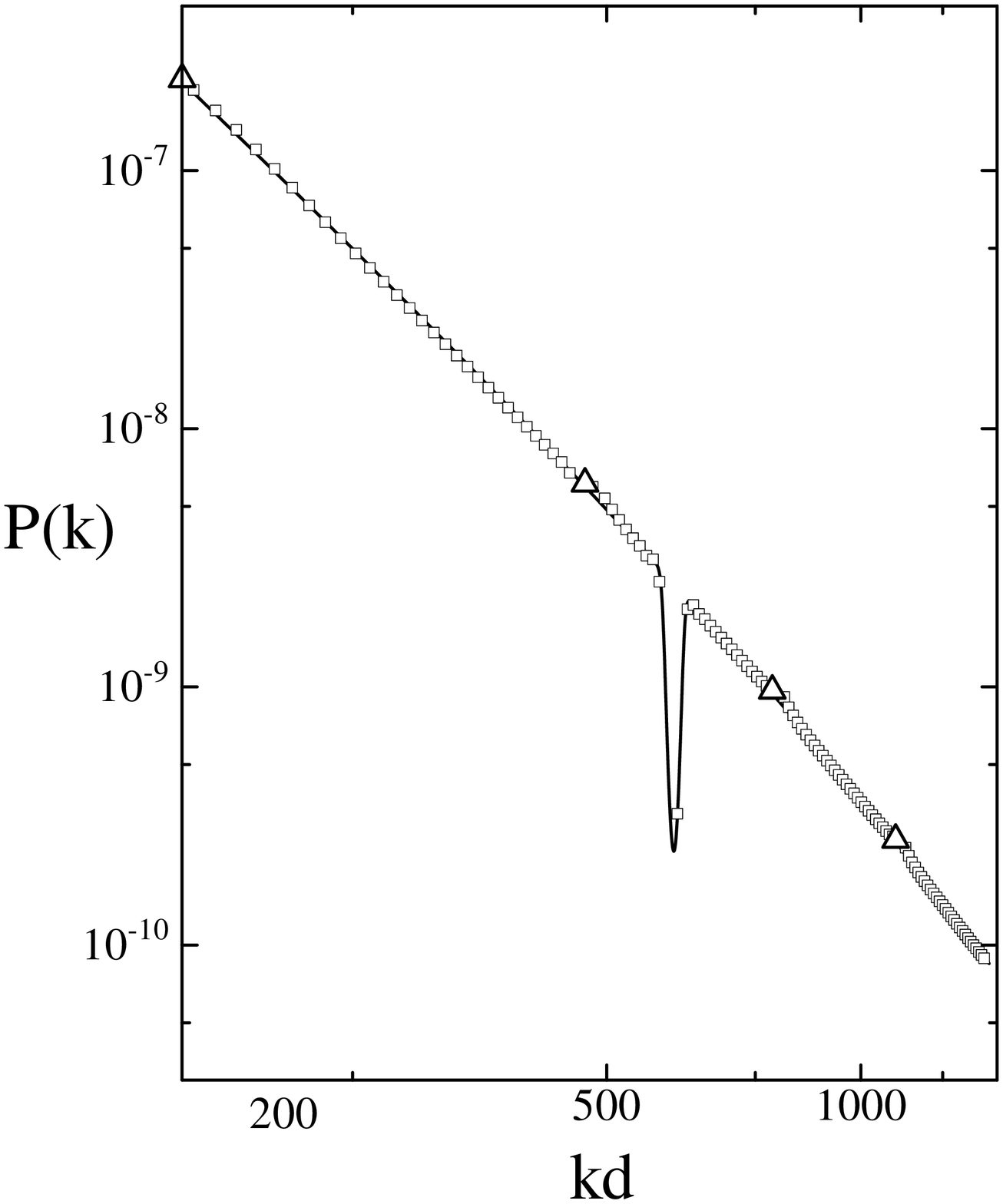}
\caption{The original spectra (the solid curves)
 and the reconstructed spectra (the boxes and triangles).
The left panel shows the case of the double
power-law spectrum given by
Eq.~(\ref{double}) and the middle and the right panels show
the single power-law spectra with a sharp peak ($s=+1$) and
a dip ($s=-1$), respectively, given by Eq.~(\ref{peaked}).
The triangles show the locations of the singular points.
We stopped the numerical integrations
in the vicinity of the singularities when
the relative error exceeded 10\,\%.}
\label{pk}
\end{center}
\end{figure}
\begin{figure}
\begin{center}
\leavevmode
\epsfxsize=6.0cm
\epsfbox{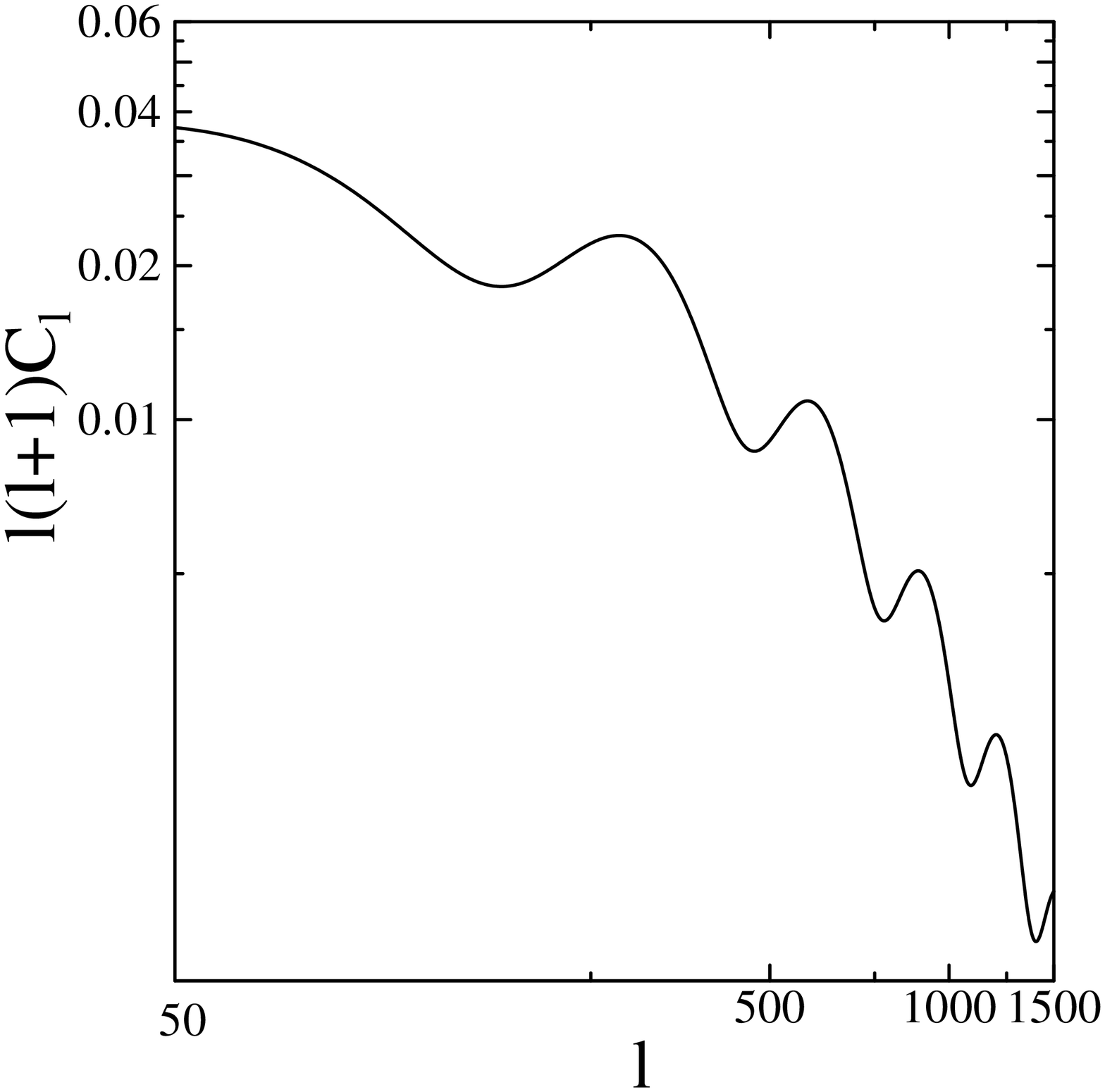}
 \hspace*{1.0cm}
\epsfxsize=6.0cm
\epsfbox{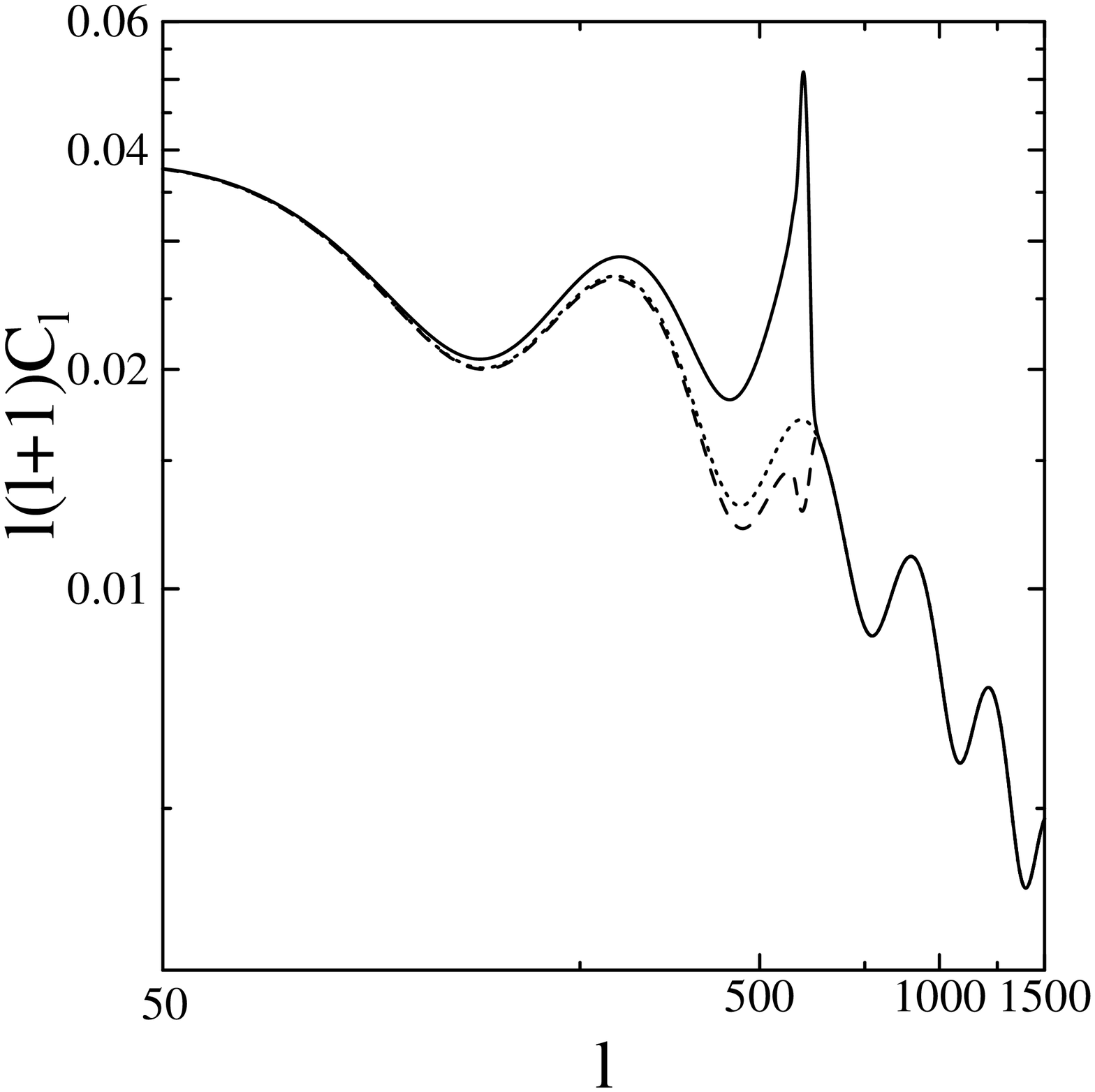}
\caption{The CMB angular power spectra for
the two initial spectra, Eq.~(\ref{double})
(left panel) and Eq.~(\ref{peaked}) (right panel).
The solid line and the dashed line on the right panel show the
case with a peak ($s=+1$) and the case with a dip ($s=-1$), respectively,
and the dotted line shows the case without the spiky structure
in the initial spectrum.
The normalization is arbitrary. Since we have neglected the
contribution of baryons to the transfer functions,
the acoustic peaks appear less prominent and their locations
are shifted to larger $l$. For realistic models with baryons,
the acoustic peaks will become more prominent and their locations
will shift to smaller $l$.
For example, the first peak, which is located at $l \sim 300$ in the
 figures, will shift to $l \sim 200$.}
\label{clspectrum}
\end{center}
\end{figure}

Finally, it is worthwhile to comment that the presence
of the singularities in the differential equation (\ref{eqdif})
may be regarded as an advantage, since the values of
$P(k)$ at the singularities can be estimated without solving
the differential equation. In particular, if there is a good reason
to believe that the spectrum should be a smoothly varying function,
a qualitative feature of the spectrum can be obtained at once.
For example, in the case of the double power-law spectrum (\ref{double}),
one can see that the original spectrum can be
approximately recovered by simply interpolating between the
adjacent triangles shown in Fig.~\ref{pk}.

\section{Conclusion}
We have considered the problem of reconstructing the initial power
spectrum of metric perturbation $P(k)$ from
$C_l$ data.
As a first step, we have investigated a simple case, namely,
the Einstein-de Sitter universe with negligible baryons and
negligible thickness of the LSS.
In this toy model, the observed temperature fluctuations
are represented only by the perturbation variables on the LSS and
the ISW effect is absent.
Then, the relation between the
initial spectrum and the angular correlation function is expressed
in terms
of an integral equation. We have shown
that this equation can be transformed to a second order
differential equation for $P(k)$. We have found that by forming
an appropriate linear combination of
the angular correlation function and its derivative,
the order of the differential equation can be reduced to the first order.
The resulting equation is found to have singularities
that come from the acoustic oscillations of photons, hence
their presence is inevitable in any cosmological models not
restricted to our simple model.
 Fortunately, however, the presence of these
singularities turns out to be not only harmless but rather advantageous.
At the singularities the coefficient of $P^{\prime}(k)$ vanishes,
and the values of $P(k)$ there can be obtained without solving
the differential equation. By plotting the values of $P(k)$, one can
obtain a rough estimate of the behavior of the spectrum.
Then, to recover the precise features of the spectrum,
the rest part of $P(k)$ can be solved with its values at
the singularities as the initial values. We have found our method
can reproduce the original spectrum with a good accuracy even for
a spectrum with a sharp, spiky structure.

The method presented here is applicable only to the Einstein-de Sitter
universe in which the ISW effect is negligible.
The ISW effect gives an important contribution even in a flat universe
model if the cosmological constant is present. Such a model, called
the $\Lambda$CDM model is preferred by recent observations\cite{LambdaCDM}.
Our next step is to include the ISW effect in our method which is
currently under study.
\acknowledgements
This work was supported in part by the Monbukagakusho/JSPS Grant-in-Aid for
Scientific Research Nos.\ 12640269(MS) and 13640285(JY),
by the Monbukagakusho/JSPS Grant-in-Aid for Priority Area: Supersymmetry and
Unified Theory of Elementary Particles (No.\ 707)(JY),
and by the Yamada Science Foundation.


\begin{thebibliography}{99}
\bibitem{WSS1}
M.~J.~White, D.~Scott and J.~Silk,
Ann.\ Rev.\ Astron.\ Astrophys.\  {\bf 32}, 319 (1994);
W. Hu, N. Sugiyama and J. Silk, Nature {\bf 386}, 37 (1997).

\bibitem{PY}
P. J. E. Peebles and J. T. Yu, Astrophys. J. {\bf 162}, 815 (1970).

\bibitem{HS}
W. Hu and N. Sugiyama, Astrophys. J. {\bf 444}, 489 (1995);
W. Hu and N. Sugiyama, Phys. Rev. D {\bf 51}, 2599 (1995).

\bibitem{power}
J. R. Bond {\it et al.} Phys. Rev. Lett. {\bf 72}, 13 (1994);
L. Knox, Phys. Rev. D {\bf 52}, 4307 (1995);
G. Jungman, M. Kamionkowski, A. Kosowski and D. N. Spergel,
 Phys. Rev. D {\bf 54}, 1332 (1995);
J. R. Bond, G. Efstathiou and M. Tegmark,
Mon. Not. R. Astron. Soc. {\bf 291}, L33 (1997);
D. J. Eisenstein, W. Hu and M. Tegmark, Astrophys. J. {\bf 518}, 2 (1999).

\bibitem{REV}
For a review of inflation, see, {\it e.g.},
A. D. Linde, {\it Particle Physics and Inflationary Cosmology}
(Harwood,
 Chur, Switzerland, 1990); K. A. Olive, Phys. Rep. {\bf 190}, 181 (1990);
 D. H. Lyth and A. Riotto, Phys. Rep. {\bf 314}, 1 (1999).

\bibitem{PER}
S. W. Hawking, Phys. Lett. {\bf 115B}, 295 (1982); A. A. Starobinsky, {\it
 ibid } {\bf 117B}, 175 (1982); A. H. Guth and S-Y. Pi,
 Phys. Rev. Lett. {\bf 49}, 1110 (1982).

\bibitem{BSI}
A. D. Linde, Phys. Lett. {\bf 158B}, 375 (1985);
L. A. Kofman and A. D. Linde, Nucl. Phys. {\bf B282}, 555 (1987);
J. Silk and M. S. Turner, Phys. Rev. D {\bf 35}, 419 (1987);
H. M. Hodges and G. R. Blumenthal, Phys. Rev. D. {\bf 42}, 3329 (1990);
J. Yokoyama and Y. Suto, Astrophys. J. {\bf 379}, 427 (1991);
M. Sasaki and J. Yokoyama, Phys. Rev. {\bf744}, 970 (1991);
A.~A.~Starobinsky,
JETP Lett.\  {\bf 55}, 489 (1992);
[Pisma Zh.\ Eksp.\ Teor.\ Fiz.\  {\bf 55} (1992) 477];
J. Yokoyama, Astron. Astrophys. {\bf 318}, 673 (1997);
J. Yokoyama, Phys. Rev. D {\bf 58}, 083510 (1998);{\bf 59}, 107303 (1999);
S.~M.~Leach and A.~R.~Liddle,
Phys.\ Rev.\ D {\bf 63}, 043508 (2001)
[arXiv:astro-ph/0010082];
S.~M.~Leach, M.~Sasaki, D.~Wands and A.~R.~Liddle,
Phys.\ Rev.\ D {\bf 64}, 023512 (2001)
[arXiv:astro-ph/0101406].

\bibitem{SOU}
T. Souradeep et. al., astro-ph/9802262;
Y. Wang, D. N. Spergel and M. A. Strauss, astro-ph/9812291;
Y. Wang, D. N. Spergel and M. A. Strauss, Astrophys. J.
{\bf 510}, 20 (1999);
S. Hannestad, Phys. Rev. D {\bf 63}, 043009 (2001).

\bibitem{BER}
A. Berera and P. A. Martin, Inverse Problems {\bf 15}, 1393 (1999).

\bibitem{SW}
R. K. Sachs and A. M. Wolfe, Astrophys. J. {\bf 147}, 73 (1967).

\bibitem{KS}
H. Kodama and M. Sasaki, Prog. Theor. Phys. Supp. {\bf 78}, 1 (1984);
H.~Kodama and M.~Sasaki,
Int.\ J.\ Mod.\ Phys.\ A {\bf 1}, 265 (1986);
H.~Kodama and M.~Sasaki,
Int.\ J.\ Mod.\ Phys.\ A {\bf 2}, 491 (1987).

\bibitem{SI}
J. Silk, Astrophys. J., {\bf 151} 459 (1968).

\bibitem{LambdaCDM}
A.~G.~Riess {\it et al.}  [Supernova Search Team Collaboration],
Astron.\ J.\  {\bf 116}, 1009 (1998)
[arXiv:astro-ph/9805201];
S.~Perlmutter {\it et al.}  [Supernova Cosmology Project Collaboration],
Astrophys.\ J.\  {\bf 517}, 565 (1999)
[arXiv:astro-ph/9812133].

\end{thebibliography}
\end{document}